\documentclass[amssymb,prl,twocolumn,floats,amsmath,showpacs,superscriptaddress]{revtex4}
\usepackage{bm}
\usepackage{graphicx}

\begin{document}


\title{Controlled aggregation of magnetic ions in a semiconductor.\\
Experimental demonstration}
%
%

\author{A.~Bonanni}\email{alberta.bonanni@jku.at}
\affiliation{Institute for Semiconductor and Solid State Physics, Johannes Kepler University, A-4040, Linz, Austria}

\author{A.~Navarro-Quezada}
\affiliation{Institute for Semiconductor and Solid State Physics, Johannes Kepler University, A-4040, Linz, Austria}

\author{Tian Li}
\affiliation{Institute for Semiconductor and Solid State Physics, Johannes Kepler University, A-4040, Linz, Austria}

\author{M.~Wegscheider}
\affiliation{Institute for Semiconductor and Solid State Physics,
Johannes Kepler University, A-4040, Linz,
Austria}

\author{ Z.~Mat{\v e}j}
\affiliation{Department of Condensed Matter Physics, Charles University, Cz-121 16 Prague, Czech Republic}

\author{V.~Hol\'y}
\affiliation{Department of Condensed Matter Physics, Charles University, Cz-121 16 Prague, Czech Republic}

\author{R.T.~Lechner}
\affiliation{Institute for Semiconductor and Solid State Physics,
Johannes Kepler University, A-4040, Linz,
Austria}

\author{G.~Bauer}
\affiliation{Institute for Semiconductor and Solid State Physics,
Johannes Kepler University, A-4040, Linz,
Austria}

\author{ M.~Kiecana}
\affiliation{Institute of Physics, Polish Academy of Sciences, PL-02-668 Warszawa, Poland}

\author{ M.~Sawicki}
\affiliation{Institute of Physics, Polish Academy of Sciences, PL-02-668 Warszawa, Poland}

\author{ T.~Dietl}
\affiliation{Institute of Physics, Polish Academy of Sciences, PL-02-668 Warszawa, Poland}
\affiliation{Institute of Theoretical Physics, University of Warsaw, PL-00-681 Warszawa, Poland}

%
\begin{abstract}
The control on the distribution of magnetic ions into a semiconducting host is crucial for the functionality of magnetically doped semiconductors. Through a structural analysis at the nanoscale, we give experimental evidence that the aggregation of Fe ions in (Ga,Fe)N and consequently the magnetic response of the material are affected by growth rate and co-doping with shallow impurities.
\end{abstract}

\pacs{75.50.Pp, 61.46.+w, 64.75.+g, 81.15.Gh}
\maketitle

There is an increasing amount of evidence that owing to specific features of magnetic impurities in wide band-gap
semiconductors and oxides, the epitaxial growth of these systems can result in the self-organized aggregation of
magnetically robust nanocrystals embedded in the host paramagnetic matrix \cite{Bonanni:2007_b,Katayama-Yoshida:2007_a,Jamet:2006_a}.
With no doubt this finding holds enormous potential for the fabrication of a range of multifunctional nanosystems
relevant to spintronics, nanoelectronics, photonics, and plasmonics \cite{Katayama-Yoshida:2007_a,Dietl:2008_a}.
However, it has also  been realized \cite{Kuroda:2007_a} that enduring difficulties in the experimental resolution
and identification of the embedded nanostructures hamper the progress in the visualization, understanding,
and control of the mechanisms accounting for relevant and hitherto unexplored nano-assembly processes.\\
\indent In this Letter, by exploiting state-of-the-art nano-characterization tools, we show how growth conditions
and co-doping with shallow donors or acceptors affect the distribution of the magnetic ions in (Ga,Fe)N deposited
by metalorganic vapor phase epitaxy (MOVPE), emphasizing conclusions that are pertinent to the whole class of wide
band-gap diluted magnetic semiconductors (DMS) and diluted magnetic oxides (DMO), extensively studied over the last
eight years \cite{Bonanni:2007_b,Malguth:2008_a,Liu:2005_a}. In particular, we identify different ways by which
transition metal (TM) impurities can incorporate into the host semiconducting lattice and we link the structure
investigated at the nano-scale to the macroscopic magnetic properties. Our findings show that Fermi-level tuning
by co-doping with shallow impurities is instrumental in controlling the magnetic ions aggregation. This provides
an experimental support to recent theoretical suggestions \cite{Dietl:2006_a,Ye:2006_a} and it demonstrates that
the Fermi-level engineering documented so-far for iodine and nitrogen doped (Zn,Cr)Te \cite{Kuroda:2007_a} operates
also in the case of III-V DMS. Moreover, we find that the aggregation of magnetic ions can be hindered by increasing
the growth rate. This indicates that the nano-assembling process is controlled by a kinetic barrier to the surface
diffusion of the magnetic ions, whose height depends on co-doping.\\
\indent All studied epilayers have been fabricated by MOVPE on $c$-plane
sapphire substrates employing the growth protocols and the {\em in
situ} and {\em ex situ} characterization methods we have reported
previously for (Ga,Fe)N \cite{Bonanni:2007_a,Pacuski:2008_a} and
GaN:Mg \cite{Simbrunner:2007_a}, and that we apply now to produce
(Ga,Fe)N and (Ga,Fe)N:Si,Mg layers. The total Fe concentration in the samples varies from
4$\times$10$^{19}$~cm$^{-3}$ to 3$\times$10$^{20}$~cm$^{-3}$ for the Fe-precursor
(Cp$_{2}$Fe) flow-rate ranging from 50 to 300
standard cubic centimeters per minute (sccm)
\cite{Bonanni:2007_a}. The Mg-concentration is found to be in the
range 2--3$\times$10$^{19}$~cm$^{-3}$ \cite{Simbrunner:2007_a}
and the Si-content is estimated to be
1$\times$10$^{19}$~cm$^{-3}$. The growth-rate during the
deposition is regulated by the Ga-precursor (TMGa) flow-rate and
varies about linearly from 0.2~to~0.3~nm~s$^{-1}$ for 5 to 15~sccm
of TMGa flow, respectively.\\
\begin{figure}
\includegraphics[width=\linewidth,clip]{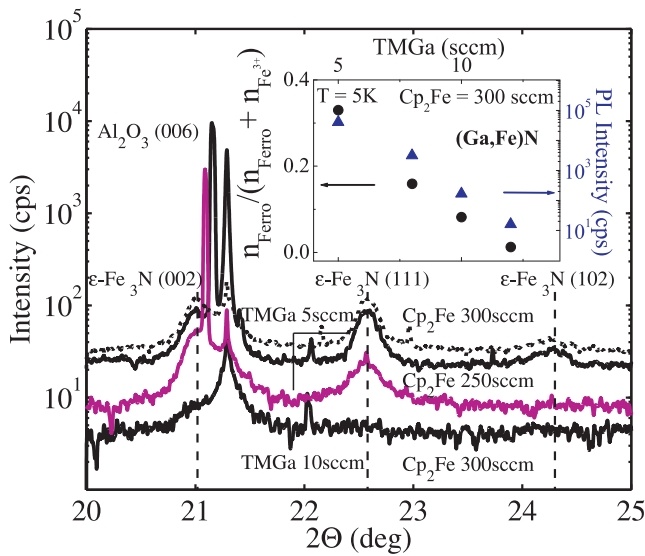}
\caption{(color online) Synchrotron radiation powder XRD spectra \textit{vs.} Fe content and growth rate for (Ga,Fe)N, as indicated. The two uppermost traces are generated by the same sample, but measured at different rotation angles (therefore different tilt). Inset: normalized number of Fe ions contributing to the ferromagnetic signatures (dots) and near-band-edge PL intensity (triangles) $vs.$ TMGa flow-rate, $i.e.$ growth-rate.}
\label{f1}
\end{figure}
\indent The difference between  the magnetization values
measured by SQUID magnetometery up to 5~T at 1.8~K and 5~K is
employed to determine the concentration $n_{\text{Fe$^{3+}$}}$ of
paramagnetic Fe$^{3+}$ ions in the layers \cite{Pacuski:2008_a},
which agrees within a factor of two with the value of
$n_{\text{Fe$^{3+}$}}$ obtained from electron spin resonance
measurements \cite{Bonanni:2007_a}. Low-field and low-temperature
hysteresis loops allow us to evaluate the concentration of Fe
contributing to the ferromagnetic signatures, $n_{\text{Ferro}}$,
which we determine assuming a magnetic moment of $2\mu_B$ per
Fe ion.\\ 
\indent In addition to high-resolution transmission electron microscopy (HRTEM) and energy dispersive spectroscopy (EDS)
\cite{Bonanni:2007_a}, we have performed and present here Fourier-filtering of the TEM images
\cite{Huang:2005_a} for strain analysis \cite{Litvinov:2006_a}, reconstructed by using the 002 spatial frequency. Powder x-ray diffraction (XRD) measurements
\cite{Stangl:2004_a}, have been carried out at the beamline ID31 of the ESRF (Grenoble -- France) using a photon energy of 15.5~keV. The x-ray data correspond to the diffracted intensities in reciprocal space
along the sample surface normals, collected using a secondary
crystal monochromator in front of the detector, thus the GaK$_{\alpha}$
fluorescence is suppressed. 
We have acquired symmetric $\omega/2\Theta$ scans for all samples with scattering angles $2\Theta$ up to 150~deg and found that the most intense peaks of the detected crystallographic phases appear in the
$2\Theta$-range up to 25~deg. As at the powder diffraction beamline the sample tilt
can not be adjusted, neither the position nor
the intensity of the substrate peaks
may be compared for different scans (see the difference between the two 
upmost curves in Fig.~1). In contrast,
the peaks for small nanocrystals are broad,
and thus virtually insensitive to the sample tilt.\\ 
\indent It has already been established that once the solubility limit of Fe in GaN at the given growth conditions is exceeded, the ferromagnetic response of (Ga,Fe)N increases with the Cp$_{2}$Fe flow-rate \cite{Bonanni:2007_a}. In Fig.~1 we report the $\omega/2\Theta$ powder diffraction
synchrotron XRD scans for (Ga,Fe)N samples with different
Fe-content above its solubility limit into the GaN host
(Cp$_{2}$Fe 250~sccm and 300~sccm,
respectively). For the samples deposited at the low growth-rate (TMGa at
5~sccm), in contrast to laboratory high-resolution XRD which does not
evidence any phase separation in (Ga,Fe)N \cite{Bonanni:2007_a}, the
high intensity available at the synchrotron beamline allows to
reveal the presence of new diffraction peaks identified as the
(002) and (111) of the phase $\varepsilon$-Fe$_3$N (hexagonal
siderazot structure, space group No.~182 (P6$_3$22), lattice
parameters $a$~=~0.4698~nm and $c$~=~0.4379~nm, Curie temperature $T_{C}$~=~575~K and magnetic moment per Fe ion $m$~=~2$\mu_{B}$ \cite{Leineweber:1999_a}). This assignment is
consistent with the crystallographic characteristics of the
Fe-rich secondary phases as put on view by TEM images, and summarized
in Fig.~2. The height of the
$\varepsilon$-Fe$_3$N-related diffraction maxima is enhanced with
increasing the flow-rate of Cp$_2$Fe, however no significant
dependence of their full-width-at-half-maxima (FWHM) is observed
upon varying the nominal content of magnetic ions. This implies that
the mean size of the precipitates does not substantially
vary with increasing nominal Fe content, whereas the density of
the precipitates is enhanced, as confirmed by TEM
\cite{Bonanni:2007_a}. We are, thus, led to the conclusion that the precipitation occurs by a nucleation mechanism in which only nanocrystals with a critical size can form.  We use the FWHM of all
diffraction maxima of $\varepsilon$-Fe$_3$N for an estimate of
the mean size of the precipitates and by employing the
Williamson-Hall plot method \cite{Warren:1990_a} we obtain an
average value for the nanocrystals diameter of $15\pm 5$~nm.\\
\indent Our studies for various growth rates demostrate that
the nanocrystal nucleation is limited by a kinetic barrier
for the surface Fe diffusion. As shown in Fig.~1 the
diffraction peaks originating from the Fe-rich nanocrystals
appear only in the slow growth-rate regime
(samples grown with TMGa at 5 sccm)
and are quenched at higher growth-rates, as evidenced by the lowest XRD trace in Fig.~1, obtained for the layer deposited at 10~sccm TMGa. The
inset of Fig.~1 gives the normalized number of Fe-ions
contributing to the ferromagnetic signatures, $n_{\text{Ferro}}$,
as a function of the TMGa precursor flow-rate ($i.e.$ the growth-rate) and its decrease with increasing
flow-rate agrees with the XRD data, giving evidence of a reduced
contribution of Fe-rich magnetic nanocrystals at higher
growth-rates. This is further supported by a dramatic drop of the
near-band-edge (3.49~eV at 10~K) photoluminescence (PL) intensity,
also reported in the inset to Fig.~1, witnessing, with increasing
growth-rate, the enhanced incorporation of the magnetic ions in
unpaired positions, where they act as PL inhibitors
\cite{Malguth:2008_a}.\\
\begin{figure*}[t]
\includegraphics[width=\linewidth,clip]{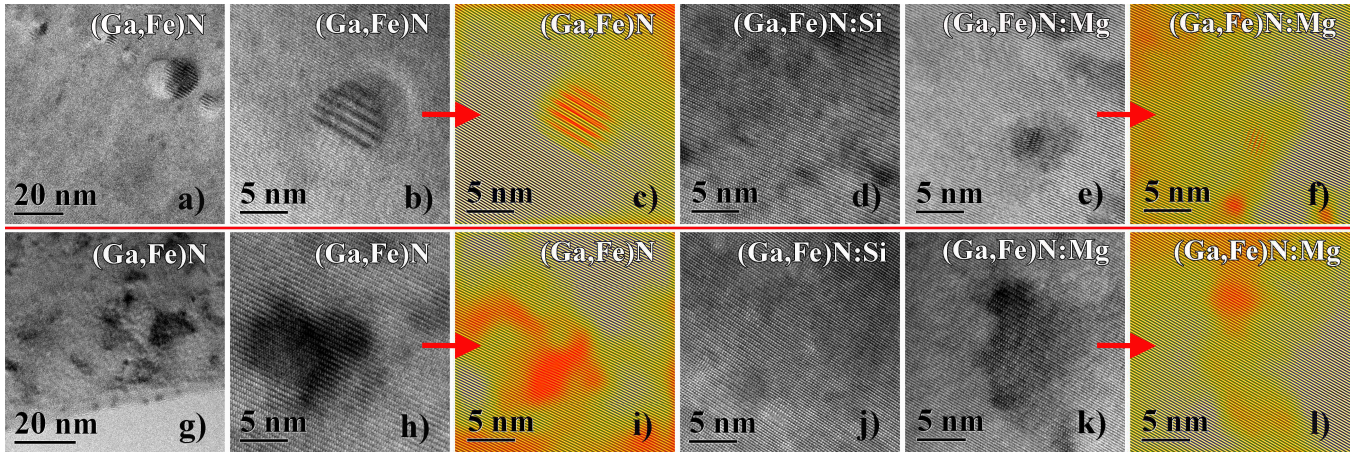} \caption{(color online) Bright-field images (a, g), HRTEM with mass contrast (b, d, e, h, j, k) and Fourier filtered images with strain mapping (c, f, i, l) of (Ga,Fe)N revealing the presence of Fe$_3$N precipitates (a, b, c, e, f), spinodal decomposition (g--l), and the effect of co-doping by either Si (d, j) or Mg (e, f, k, l) preventing the formation of the Fe-rich regions.}
\label{f2}
\end{figure*}
\indent At intermediate values of growth rate ($e.g.$ TMGa at 10 sccm) and a Fe-precursor flow-rate of 300~sccm, where with synchrotron XRD the onset of second phases is not detected, ferromagnetic signatures are still clearly seen by SQUID up to a blocking temperature typically over 300~K. These puzzling observations are elucidated by the TEM images with mass contrast and strain mapping, shown in Figs.~2(g--l), that put into evidence the presence of spinodal decomposition into nano-scale regions with high Fe content embedded in the Fe-poor matrix without any crystallographic phase separation. This appears as a generic property of a number of DMS and DMO, in which no precipitates are detected, but ferromagnetic features persist up to high temperatures \cite{Liu:2005_a}.\\
\indent Remarkably, our TEM, XRD, and SQUID data reveal that the aggregation of Fe ions can be diminished or even prevented by co-doping with either Si donors or Mg acceptors. Figure~2 presents TEM data for the two relevant initial regimes, namely (Ga,Fe)N with embedded Fe-rich nanocrystals [Figs.~2(a--f)] evidenced by Moir$\acute{e}$ fringes contrast and (Ga,Fe)N showing spinodal decomposition [Figs.~2(g--l)] generating mass contrast and lattice distortion, as proved by the filtered images [Figs.~2(i, l)]. From Figs.~2(d--f) and 2(j--l) the reduced aggregation of Fe-rich regions as a consequence of co-doping with Si and Mg, respectively, is evident.

\begin{figure}[b]
\includegraphics[width=\linewidth,clip]{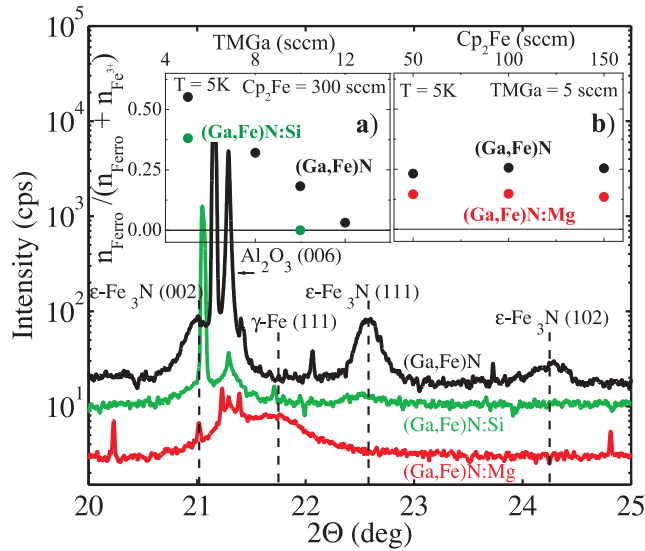} \caption{(color online) Effect of Si and Mg co-doping on the synchrotron radiation powder XRD spectra of (Ga,Fe)N; insets: effect of co-doping (a) with Si and (b) with Mg on the normalized number of Fe ions contributing to the ferromagnetic signatures in (Ga,Fe)N \textit{vs.}~TMGa flow-rate, \textit{i.\,e.} growth-rate (a) and \textit{vs.}~Cp$_{2}$Fe flow-rate (b).}
\label{f3}
\end{figure}

This effect is further corroborated by the XRD results
given in Fig.~3. In both cases, the shallow impurities are found to hamper efficiently the precipitate aggregation, so that the diffraction peaks corresponding to $\varepsilon$-Fe$_3$N are suppressed in the case of co-doped samples.  We point out that the XRD curve from the Mg co-doped sample exhibits also a broad maximum identified as the (111) diffraction from pure $\gamma$-Fe (austenite). This finding is supported by our TEM observations, revealing in the considered sample the presence of a few-nm thick Fe inclusions at the surface.
The suppression of the ferromagnetic contribution in the co-doped layers is further validated by the reduced number of average Fe ions adding to the ferromagnetic response in both Si-- and Mg--co-doped (Ga,Fe)N layers. In inset a) to Fig.~3, the normalized $n_{\text{Ferro}}$ is given for a constant Cp$_{2}$Fe as a function of the TMGa flow-rate. As seen here, the quenching of the ferromagnetic contribution by Si co-doping is equally observed in samples deposited at a low growth-rate (5~sccm TMGa, \textit{i.\,e.} presenting Fe-rich nanocrystals when not co-doped) and at intermediate growth-rate (10~sccm TMGa, \textit{i.\,e.} showing -- when not co-doped -- spinodal decomposition). The change in the normalized $n_{\text{Ferro}}$ upon co-doping with acceptors is presented in inset b) as a function of the nominal Fe content in the low magnetic ions doping regime (Cp$_{2}$Fe at 50--150 sccm).\\
\indent In order to explain these key findings, we note that except for Mn in II-VI compounds \cite{Kuroda:2007_a}, owing to the presence of the open $d$ shells in the vicinity of the Fermi level, the nearest neighbor pair of TMs in semiconductors shows a large binding energy which promotes the magnetic ions aggregation \cite{Katayama-Yoshida:2007_a,Kuroda:2007_a,Schilfgaarde:2001_a}. However, if carriers introduced by co-doping can be trapped by these ions, the pair binding energy will be altered, usually reduced by the corresponding Coulomb repulsion \cite{Kuroda:2007_a,Dietl:2006_a,Ye:2006_a}. While the presence of the mid-gap electron trap, \textit{i.\,e.,} the Fe$^{+3}$/Fe$^{+2}$ state, is well established in GaN \cite{Malguth:2008_a}, the level Fe$^{+3}$/Fe$^{+4}$ is expected to reside rather in the valence band \cite{Malguth:2008_a}. However, it has been recently suggested that in GaN the potential introduced by the Fe$^{+3}$ ion is strong enough to trap a hole in a Zhang-Rice type of state \cite{Dietl:2008_b}. Therefore, since the concentrations of Si and Mg are comparable to
that of $n_{\text{Ferro}}$ in our samples, the obstructive effect
of Si and Mg co-doping on the nanocrystal formation and, thus, on
the ferromagnetic response, is elucidated. Moreover, in order to
test the scenario that the Fe aggregation occurs at the growth
surface we deposited films in which Fe and Mg have been
alternatively supplied in the $\delta$-like fashion. No influence
of co-doping on the ferromagnetism has been found, despite that Mg
is known to diffuse in GaN \cite{Simbrunner:2007_a}.
Significantly, in view of our results, the previously observed effect of co-doping on ferromagnetism in (Ga,Mn)N, and assigned to the dependence of the double exchange mechanisms of the spin-spin coupling on the position of the Fermi level with respect to the center of the $d$ band \cite{Reed:2005_a}, has to be reconsidered. Furthermore, since with the experimental resolution of 5$\times$10$^{16}$~$\mu_{B}$~cm$^{-3}$ 
we do not observe any spontaneous magnetization in GaN, GaN:Si and GaN:Mg down to 5~K, we conclude that the recently invoked vacancy-related ferromagnetism of GaN \cite{ Dev:2008_a} is not present in our samples.\\
\indent In summary, by combining HRTEM and synchrotron XRD with SQUID we have identified three distinct ways by which Fe incorporates into the GaN lattice: (i) substitutional Fe$^{3+}$ diluted ions accounting for the paramagnetic response \cite{Bonanni:2007_a}; (ii) Fe-rich (Ga,Fe)N wurtzite nanocrystals commensurate with and stabilized by the GaN host lattice and (iii) hexagonal $\epsilon$-Fe$_3$N precipitates. 
The formation of nanocrystals containing a large density of the magnetic constituent elucidates the origin of the ferromagnetic features persisting up to above room temperature. Importantly, the co-doping with either Si or Mg hampers the nanocrystal assembling. This demonstrates that the charging of the magnetic ions, Fe$^{3+}\rightarrow$ Fe$^{2+}$ and Fe$^{3+}$ $\rightarrow$ Fe$^{3+} + h$, respectively, inhibits the Fe aggregation and explains the sensitivity of the ferromagnetic response to co-doping with shallow donors or acceptors.  Furthermore, the influence of the growth rate on the nanocrystal formation indicates that the Fe aggregation occurs at the growth surface. Since, quite generally, the binding energy of TM pairs depends on the valency of the open $d$-shells, there is a ground to suppose that the Fermi level engineering evoked here for (Ga,Fe)N:Si,Mg can serve to control the magnetic ion aggregation in a number of semiconductors and oxides, providing a way to the self-organized fabrication of multi-component systems with tailored magnetic, magneto-optical, and magneto-transport properties at the nanoscale.\\
\indent This work has been supported by the Austrian Fonds zur
{F\"{o}rderung} der wissenschaftlichen Forschung (P18942, P20065 and N107-NAN), by the Czech Grant Agency
(202/06/0025) and Ministry of Education (MSM 0021620834), and by the EC
(NANOSPIN, FP6-2002-IST-015728; SPINTRA,
ERAS-CT-2003-980409). We acknowledge the assistance of the ID31
staff at the ESRF.

\end{document}